\documentclass[12pt,preprint]{aastex}

\def\lya{Ly$\alpha$}

\def\ergcm2s{\ifmmode {\rm\,erg\,cm^{-2}\,s^{-1}}\else
                ${\rm\,ergs\,cm^{-2}\,s^{-1}}$\fi}
\def\ergsec{\ifmmode {\rm\,erg\,s^{-1}}\else
                ${\rm\,ergs\,s^{-1}}$\fi}
\newcommand{\fluxunit}{\ergcm2s}

\begin{document}

\title{X-ray properties of the $z$ $\sim$ 4.5 \lya\ Emitters in the Chandra Deep Field South Region}%: Possible weak AGN in High-z Star-Forming Galaxies}

\author{Z. Y. Zheng\altaffilmark{1, 3}, J. X. Wang\altaffilmark{1}, S. L. Finkelstein\altaffilmark{2},
S. Malhotra\altaffilmark{3}, J. E. Rhoads\altaffilmark{3}, K. D. Finkelstein\altaffilmark{2}}

\begin{abstract}
%We find that X-ray observation is very powerful to check AGN and exclude interlopers 
%in  
  We report the first X-ray detection of \lya\ emitters at redshift
  z$\sim$4.5. One source (J033127.2-274247) is detected in the
  Extended Chandra Deep Field South (ECDF-S) X-ray data, and has been
  spectroscopically confirmed as a $z = 4.48$ quasar with $L_X =
  4.2\times 10^{44} \ergsec$. The single detection gives a \lya\ quasar density of
  $\sim$ 2.7$^{+6.2}_{-2.2}\times$10$^{-6}$ Mpc$^{-3}$, consistent with the X-ray
  luminosity function of quasars.  Another 22 \lya\ emitters (LAEs) in the central Chandra
  Deep Field South (CDF-S) region are not detected individually, but their
  coadded counts yields a S/N=2.4 (p=99.83\%) detection at soft band, %of
  %$F_{0.5-2.0 keV}$ = 8.8$\pm$3.7 $\times 10^{-18}$ \fluxunit
  with an effective exposure time of $\sim 36$ Ms. 
  Further analysis of the equivalent width (EW) distribution shows that all the signal comes from 
12 LAE candidates with EW$_{rest} <$ 400 \AA, and 2 of them contribute about half of the signal. 
From follow-up spectroscopic observations, we find that one of the two is a low-redshift emission line galaxy, and the other is a Lyman break galaxy at z = 4.4 with little or no Ly$\alpha$ emission.
 Excluding these two and combined with ECDF-S data, we derive a 3-$\sigma$ upper limit on the 
 average X--ray flux of $F_{0.5-2.0 keV}$ $<$ 1.6 $\times
  10^{-18}$ \fluxunit,  which corresponds to an average luminosity of
  $\langle L_{0.5-2 keV} \rangle$ $<$ 2.4 $\times 10^{42}$ ergs $s^{-1}$ for z $\sim$
  4.5 \lya\ emitters.  If the average X-ray emission is due to star formation, it corresponds to a
  star-formation rate (SFR) of $<$ 180--530 M$_\sun$ yr$^{-1}$. We use this SFR$_X$ as an upper limit of the
  unobscured SFR to constrain the escape fraction of \lya\ photons, and find a lower limit of f$_{esc,Ly\alpha}$ $>$ 3--10\%.
  However, our upper limit on the SFR$_X$ is $\sim$7 times larger than the 
  upper limit on SFR$_X$ on z$\sim$ 3.1 LAEs in the same field, and at least 30 times
  higher than the SFR estimated from \lya\ emission.
 % These discrepancies imply that our marginal signal is likely dominated by 
 % weak AGN. 
 From the average X-ray to \lya\ line ratio, we estimate that fewer than 3.2\% (6.3\%) of
 our LAEs could be high redshift type 1 (type 2) AGNs, and those hidden AGNs likely show low rest frame equivalent widths. 
%  X-ray observation is very powerful to resolve AGN and interlopers in the LAE sample. 
 
  %Further analysis of the equivalent width (EW) distribution shows that all the signal comes from 
 % 12 LAEs with EW$_{rest} <$ 400 \AA, and is probably dominated by weak AGN.
  
 % re exists weak AGNs in the high-z LAE sample. 

 %  The stacked X-ray color
%  (HR$<$-0.26) shows that the X-ray signal at soft band is dominated
%  by star-forming galaxies or type 1 AGNs than by type 2 AGNs.  If the
%  X-ray is due to star formation, it corresponds to an upper limit on
%  star-formation rate (SFR) 100--300 M$_\sun$ yr$^{-1}$, one decade
%  higher than the SFR estimated from \lya\ emission.  The difference
%  in SFRs can be used to constrain that at least 6\% of \lya\ photons
%  escape.

\end{abstract}


\altaffiltext{1}{Center for Astrophysics, University of Science and
Technology of China, Hefei, Anhui 230026, China; zhengzy@mail.ustc.edu.cn, 
jxw@ustc.edu.cn.} 

\altaffiltext{2}{George P. and Cynthia Woods Mitchell Institute for Fundamental Physics and Astronomy, Department of Physics and Astronomy, Texas A\&M University, College Station, TX 77843} 

\altaffiltext{3}{School of Earth and Space Exploration, Arizona State University, Tempe, AZ 85287}

\section {INTRODUCTION}

Narrowband surveys have discovered thousands of candidate \lya\
emitters from z $=$ 2.25 -- 6.96 (e.g., Nilsson et al. 2009,
Gawiser et al. 2007, Rhoads et al. 2000, 2003, Dawson et al. 2007,
Ouchi et al. 2008, Wang, et al. 2005, Iye et al. 2006).  Hundreds 
have  been spectroscopically confirmed (e.g., Hu et al. 2004,
Dawson et al. 2004, Venemans et al. 2005, Dawson et al. 2004, 2007,
Ouchi et al. 2008, Wang et al. 2009).  Recent studies have found
evidence for dust in \lya\ galaxies (e.g., Finkelstein et al. 2008,
2009c, Lai et al. 2007, Pirzkal et al. 2007), showing that \lya\
galaxies are not all primitive. This dust may help to explain the
``problem'' of the observed equivalent widths (EWs) of high-$z$ LAEs.
These EWs are often larger than expected even from normal star
formation (Malhotra \& Rhoads 2002).  Possible scenarios for causes of
these large EWs include very low metallicities, or enhancement of the
Ly$\alpha$ EW via a clumpy interstellar medium (ISM; Neufeld 1991,
Hansen \& Oh 2006; Finkelstein et al.\ 2009c).

Active Galactic Nuclei (AGNs) can also account for high \lya\ EWs of \lya\ emitters (LAEs
hereafter).  X-ray studies of LAEs can help us to detect AGN.
However, unlike the LAEs in the local universe, where the AGN fraction
is as high as 15-40\% (e.g., Scarlata et al. 2009, Cowie et
al. 2010 and Finkelstein et al. 2009a, 2009b)\footnote{Note that these
studies use methods beyond X-rays, e.g., optical emission line
diagnostics.}, the observed AGN fraction at high redshift is small,
from 3--7\% at z=2.1 (Guaita et al. 2010),  5--13\% at z$\sim$2.25
(Nilsson et al. 2009), 1--5\% at z$\sim$3.1-3.7 (Gronwall et al. 2007; Ouchi
et al. 2008; Lehmer et al. 2009), to $<$ 5\% at z$\sim$4.5 (Malhotra
et al. 2003, Wang et al. 2004) and $<$ 1\% at z$\sim$5.7 (Ouchi et
al. 2008).  This trend is in line with the observed decrease in the
number density of quasars at z $>$ 2 (e.g., figure 14 of Yencho et al. 2009).

In addition to measuring AGN contributions, X-ray emission is also a
useful measure of the unobscured star-formation activity, mainly from
 supernovae (SNe), hot interstellar gas (i.e., $T >
10^{6-7}$ K), high-mass X-ray binaries (HMXBs), and low-mass X-ray
binaries (LMXBs).  The first three object classes evolve rapidly, and
therefore track the current star-formation rate (SFR).  The LMXBs have
longer evolutionary time scales (on the order of the Hubble time), and
therefore track the integrated star-formation history of galaxies
(i.e., the total stellar mass).  Colbert et  al. (2004) give a relationship of
L$_{2-8 keV}$ = $\alpha\times$M$_*$ + $\beta\times$SFR from X-ray observations of nearby galaxies, 
where Lx , M$_*$, and SFR have units of ergs s$^{-1}$ , M$_{\odot}$ , and M$_{\odot}$ yr$^{-1}$, 
respectively, and constants $\alpha$ = 1.3$\times$10$^{29}$ ergs s$^{-1}$ M$_{\odot}^{-1}$ 
and $\beta$ = 0.7$\times$10$^{39}$ ergs s$^{-1}$ (M$_{\odot}$ yr$^{-1}$)$^{-1}$.
When SFR $>$ 5 M$_{\odot}$ yr$^{-1}$, many authors (Grimm et al. 2003,
Ranalli et al. 2003, Persic et al. 2004) show that  the galaxies' non-nuclear X-ray emission
can be used as a linear star formation rate indicator for high redshift star-forming galaxies, 
which might be dominated by HMXBs.  
Laird et al. (2005) stacked the X-ray flux from UV-selected star-forming galaxies at z$\sim$1 in the Hubble 
Deep Field North, and found a mean 2-10 keV rest-frame luminosity of 2.97$\pm$0.26 $\times$ 10$^{40}$ ergs s$^{-1}$,
corresponding to an X-ray derived SFR (hereafter SFR$_X$) 
of 6.0$\pm$0.6 M$_{\odot}$ yr$^{-1}$,  derived using the conversion from Ranalli et al. 2003.
This is $\sim$3 times the mean UV derived SFR (hereafter SFR$_{UV}$). 
In the same field, Laird et al. (2006) found the average SFR$_X$ of 42.4$\pm$7.8 M$_{\odot}$ yr$^{-1}$ for z $\sim$ 3 LBGs, 
about  4.1 times SFR$_{UV}$.  Additionally, Lehmer et al. (2005) reported the average SFR$_X$ 
of $\sim$30 M$_{\odot}$ yr$^{-1}$ for z $\sim$ 3 LBGs in the {\it Chandra} Deep Field -- South (CDF-S). Lehmer et al. also stacked LBGs in the CDF-S at z$\sim$4, 5, and 6, 
and did not obtain significant detections ($<$3 $\sigma$), deriving rest-frame 2.0-8.0 keV luminosity upper 
limits (3 $\sigma$) of 0.9, 2.8, and 7.1 $\times$ 10$^{41}$ ergs s$^{-1}$, corresponding to 
SFR$_X$ upper limit of 18, 56 and 142 M$_{\odot}$ yr$^{-1}$, respectively. Note also that a $\sim$3 $\sigma$ stacking signal of 
the optically bright subset (brightest 25\%) of LBGs at z$\sim$4 was detected, corresponding to an average SFR$_X$ 
of $\sim$28 M$_{\odot}$ yr$^{-1}$. 
These studies demonstrate the value of stacking the deepest X-ray observations
to obtain sensitive detections or strong upper limits on star formation activity, with
little sensitivity to dust.

 Since LAEs are thought to
be less massive and much younger than LBGs at high-redshift 
(e.g., Venemans et al 2005; Pirzkal et al 2007; Finkelstein et al 2008, 2009c)
their X-ray emission is probably due to the newly formed HMXBs.  An X-ray
detection could give us an unbiased SFR estimate, or more properly an
upper limit, since AGN may contribute to the X-ray flux.

The first X-ray observations of high--redshift LAEs were 
presented in Malhotra et al. (2003) and Wang et al. (2004) at z$\sim$4.5 with two 170 ks {\it Chandra} exposures.
No individual LAEs were detected, and a 3-$\sigma$ upper limit on the X--ray
 luminosity (L$_{2-8 keV}$ $<$ 2.8 $\times$ 10$^{42}$ ergs s$^{-1}$) was derived by an X-ray stacking method (Wang et al. 2004). 
From a stacking analysis of the non-detected LAEs in the 2 Ms CDF-S field, Gronwall et al. (2007) and Guaita et al. (2010) found a smaller 
3-$\sigma$ upper limit on the luminosity of $\sim$3.1 $\times$ 10$^{41}$ ergs s$^{-1}$ and 1.9 $\times$ 10$^{41}$ ergs s$^{-1}$ at z = 3.1 and z = 2.1.  These imply upper limits of unobscured SFR$_X$ $<$  70 M$_{\odot}$ yr$^{-1}$ and $<$ 43  M$_{\odot}$ yr$^{-1}$, respectively (using the L$_{X}$ - SFR calibration of Ranalli et al. 2003).
 Until now, there has been no detection of LAEs at
$z>$4 in the X-rays, even with stacking analyses (Malhotra et
al. 2003, Wang et al. 2004, Ouchi et al. 2008).  In this paper, we
match 113 z $\sim$ 4.5 LAE candidates with the deepest 2 Ms $Chandra$
exposure of the \textit{Chandra} Deep Field South (CDF-S), and a
shallower ($\sim$ 240 ks) but wider-area exposure of the Extended
\textit{Chandra} Deep Field South (ECDF-S).

\section{OPTICAL AND X-RAY DATA}

The LAE candidates were selected with narrowband imaging of the GOODS
CDF-S (RA 03:31:54.02, Dec
-27:48:31.5, J2000)  at the Blanco 4m telescope at Cerro
Tololo InterAmerican Observatory (CTIO) with the MOSAIC II camera.
Three 80 \AA\ wide narrowband filters (NB656, NB665 and NB673) were
utilized to obtain deep narrowband images (Finkelstein et al.
2008, 2009c). The LAE candidates are  selected based on a 5 $\sigma$
detection in the narrowband, a 4 $\sigma$ significant
narrowband flux excess over the broad band continuum image (here, an R band image
from the ESO Imaging Survey [EIS], Arnouts et al. 2001), a factor of 2 ratio of
narrowband flux to broadband flux density, and no more than 2
$\sigma$ significant flux in the EIS-B band. Candidates with GOODS
B-band coverage were further examined in the GOODS B-band image, and
those with significant B-band detections were excluded.
These conditions are
satisfied by 113 LAE candidates with the \textit{Chandra} CDF-S and ECDF-S coverage, 
including 4 in the NB656 filter\footnote{The NB656
data was much shallower than the other two bands, thus the galaxies were 
selected in a different way (see Finkelstein et al. 2008) - we search for the NB656 candidates from the
 positions of galaxies which were detected in GOODS V-band but not in GOODS B-band.  Thus, we were only 
 able to select galaxies over the GOODS region,
 which is why only four objects were
selected. The other two catalogs consist of all selected candidates
over the overlap region between the MOSAIC image and the ESO Imaging
Survey, which consists of a much larger area. }
(%Only in GOODS CDF-S area, 
Finkelstein et al. 2008),  39  in NB665, and 81 in NB673 (%about ECDF-S area, 
including 11 that were detected in both NB665 and NB673).  
The equivalent widths (EWs) of our LAEs were calculated from our narrowband and EIS-R broadband data.
Finkelstein et al (2008, 2009c) have previously studied the 14 objects
from this sample that lie within the GOODS \textit{HST} field. For these sources, we choose the deeper GOODS V-band
to calculate the EWs. 

The 2 Ms  \textit{Chandra X--Ray Observatory} ACIS (Advanced CCD Imaging Spectrometer) exposure of the 
CDF-S is composed of 23 individual ACIS-I observations. We downloaded the raw data
from the \textit{Chandra} public archive and reduced the data using the
\textit{Chandra}  Interactive Analysis of Observations software version 4.0
(CIAO4.0). Each observation was filtered to include only standard
\textit{ASCA} event grades 0, 2, 3, 4, 6. Cosmic ray afterglows, ACIS hot pixels, and
bad pixels were removed, along with all data taken during high background time intervals.
All exposures were then added to produce a combined  event file with a net exposure of 1.9 Ms.
The  \textit{Chandra} exposure of the ECDF-S is composed of 9 individual ACIS-I
observations obtained in 2004, covering $\sim$ 0.3 deg$^2$ with four pointings.
We reprocessed the X-ray raw data of the four pointings separately. The averaged
net exposure per pointing at ECDF-S was 238 ks. The aspect 
offset\footnote{http://cxc.harvard.edu/cal/ASPECT/fix\_offset/fix\_offset.cgi}
of both CDF-S and ECDF-S data was examined and no offset above 0.1\arcsec\ was found in
either field. We used the published X-ray source catalogs of the 2-Ms CDF-S (Luo et al. 2008) and 
the 240-ks ECDF-S (Lehmer et al. 2005) in the following source-match and source-mask processes.

\section{X-RAY IMAGING RESULTS}

\subsection{X-ray Individual Detection}
In this paper we focus on the X-ray data with an off-axis angle $ <
8 \arcmin$, because the spatial resolution of ACIS-I data degrades rapidly
for off-axis angles $> 8'$. This excludes 22 LAEs from our sample, leaving a total of 91 LAE at $z\sim4.5$ covered by {\it Chandra} images with
off-axis angle $\theta$ $<$ 8\arcmin.  Of these, 22 are covered by the 2 Ms CDF-S
exposure, and 86 by the shallower ECDF-S exposures, with 17 sources covered by both
(see Figure 1). We choose a radius of 3\arcsec\ to match X--ray counterparts to our LAE sample, as our narrowband data has a
seeing of 0.9\arcsec\, and the radius of 50\% PSF regions of Chandra ACIS-I reaches 2.8\arcsec\ at 
the edge of our selection area. 
Only one LAE (J033127.2-274247) has an individually detected X-ray counterpart 
(ECDFS-J033127.2-274247), with a spatial offset $\le$ 0.4\arcsec\ between the NB673 
and X-ray coordinates.  This object has previously been spectroscopically identified 
as an unobscured z $=$ 4.48 quasar (Treister et al. 2009), with full--band luminosity of
L$_{0.5-10 keV}$= 4.2 $\times$ 10$^{44}$ ergs s$^{-1}$ (assuming
$\Gamma$ = 1.4, Lehmer et al. 2005).
We measured the $f_{Ly \alpha}/f_{0.5-10 keV} \approx 0.065$,
consistent with expectations from a quasar template 
($f_{Ly \alpha}/f_{0.5-10 keV}$ $\sim$ 0.05, Sazonov et al. 2004).

There are two LAEs (J033204.9-280414 and J033154.1-274159)
located in 95\% PSF circles of two ECDF-S sources with offsets between
X-ray and optical of 4.5\arcsec\ and 4.1\arcsec, respectively.  These
 offsets are too large to reliably associate the \lya\ and X-ray sources, thus we
do not classify them as X-ray detections, and we exclude these sources
from our X-ray stack (Sec.~\ref{sec:stacka}).

We plot the X-ray signal-to-noise ratio distribution of the remaining X-ray flux
measurements in Figure~\ref{stacksn}.  This comprises 106 exposures on 88 distinct LAEs (22 are covered by the 2 Ms CDF-S exposure, and 84 by the shallower ECDF-S exposures, with 17 sources covered by both 
and one by two ECDF-S pointings, see Figure 1).
The S/N ratios were calculated as $S/N = S/(\sqrt{T+0.75}+1)$  (Gehrels 1986), 
where $S$ and $T$ are the net counts and total counts extracted from their 50\% PSF
circles\footnote{
The background counts (${\rm B = T - S}$) were first extracted from an annulus with
1.2R$_{95\%PSF}$ $<$ R $<$ 2.4R$_{95\%PSF}$ (after masking out nearby
X-ray sources), and then scaled to their 50\% PSF regions by dividing the 
ratio of cumulated exposure in background region and source region. } at 0.5-2, 2-7 and 0.5-7 keV bands,
respectively. When converting from PSF-corrected count-rate  to
flux, the full and hard bands were extrapolated to the standard
upper limit of 10 keV.  All X-ray fluxes have been corrected for Galactic 
absorption (Dickey \& Lockman 1990).  
To convert from X-ray counts to fluxes, we have assumed powerlaw spectra with photon index of 
$\Gamma$ = 2 (except where explicitly stated otherwise), which generally represents the X-ray spectra of both starburst galaxies and type 1 AGNs.
For the LAEs without individual X-ray detections, we derived $3-\sigma$ upper 
limits to their X-ray fluxes (see Figure \ref{fslya}).

\subsection{Stacking analysis} \label{sec:stacka}
To determine the mean X-ray properties of the high--redshift LAEs that are
too weak to be directly detected, we employed a stacking technique
similar to that described in Wang et al. (2004) and Laird et al.
(2006). The only difference was on the count-extraction, we chose the 50\% PSF regions here 
for the nondetections rather than 80\% PSFs (as in Wang et al) or fixed radius
of 1.5$\arcsec$ (as in Laird et al).  Small apertures give better upper limits 
on non-detections, and constant size is difficult for flux estimation here.  After masking out the detected X-ray sources, the net and
background counts were measured in the CDF-S and ECDF-S separately (see Figure 1 and Table 1).
We computed two stacks:    (a) The CDF-S data alone, and (b) all available data (CDF-S and
ECDF-S).  For objects in the overlap between the CDF-S and ECDF-S coverage, only their CDF-S
data was included in stack (a), while data from both images was included in stack (b).

A marginal signal was found from the stacking of the CDF-S data in the soft band, while no signals were found in the other band 
of CDF-S or from the ECDF-S stacking.
%No signals were found from the ECDF-S stacking, while a marginal signal was found from the stacking of the soft band CDF-S data.  
When cumulating  the 22 LAEs in the central CDF-S in the soft X-ray band,
we measure net and background counts of 26.6 and 74.4, which yields a signal--to--noise (S/N) =  2.4, with an effective exposure
time of $\sim$ 36 Ms. The net counts and exposure time can be converted to an average flux of 
$F_{0.5-2.0 keV}$ = 8.8$\pm$3.7 $\times 10^{-18}$ \fluxunit.
Including the ECDF-S data, the net exposure rises to 52 Msec for 106 exposure of 88 objects, while the 
net and background counts only increased 0.2 and 26 at soft band, respectively, corresponding to a slightly
decreased S/N of 2.2.   In principle, the increase in effective exposure of a factor 1.44 (52Ms/36Ms) 
should imply an expected value of S/N = 2.4$\times \sqrt{1.44}$ = 2.9. However,  given the small
numbers of X-ray photons involved, the count rates in the CDF-S and ECDF-S are in fact
consistent at $<2\sigma$.  The effective 52 Msec exposure decreases our soft band signal to
%. This might due to the not
%deep enough exposure of ECDF-S compared with CDF-S, since we got an average of background 
%counts of 3.4 per LAE in CDF-S at soft band, compared to  0.3 in ECDF-S. 
%So not strange the S/N goes down to S/N = 2.2, and
% the soft band signal decreases to 
$\langle f_{0.5-2}\rangle = 6.4\pm 2.9 \times
10^{-18}$ \fluxunit, corresponding to an average 
luminosity of $\langle L_{0.5 - 2 keV}\rangle$ = 1.3 $\times 10^{42}$ ergs $s^{-1}$ for LAEs at z $\sim$ 4.5, 
%We obtained net and background counts of 27 and 100 in the soft
%band, and an effective exposure time of 52 Ms (of which 35.8 Ms
%comes from the CDF-S, and 16.2 Ms from ECDF-S, see Table~\ref{stkdata}). 
%This soft band signal implied an average flux of $\langle f_{0.5-2}\rangle = 6.4\pm 2.9 \times
%10^{-18}$ \fluxunit\ for LAEs at z $\sim$ 4.5. 
We also stacked the images of our LAEs together.  Since the stacked samples in
ECDFS were within their background fluctuations, we only stacked the
22 LAEs located in CDF-S.  The resulting stacked image shows a signal consistent
with the analysis above, which becomes apparent to visual inspection when
smoothed with a Gaussian matched to the ACIS PSF size (see Figure \ref{stack}).

We performed Monte Carlo simulations to check the significance of the stacked signal. 
By randomly choosing 22 positions on the source-masked CDF-S image, then cumulating 
source and background counts, we obtained distributions of both the
net counts and the soft band S/N distribution 
(Figure \ref{stacksn}).   The net counts in the simulations agreed very 
well with a Poisson distribution having a mean of 74 (the expected total
background counts in 22 apertures).  Both the Monte Carlo simulation and
the Poisson distribution gave a probability of P(S/N$<$2.4) =
99.83\% for obtaining a signal as strong as the observed one by chance.  
We also use a {\emph{jackknife}} test on our stacking result (see
insets in fig.~\ref{stacksn}).  This test is to validate the sample by
using subsets of the data from which one or two sources have been
excluded.  The jackknife test shows that there are two sources that
contribute about half of the stacked signal.  We regard these as
suspected X-ray sources.%, but given the statistical uncertainties we
%cannot say with any certainty whether our stacked signal is truly
%dominated by these two sources or is instead dominated by a lower
%level of emission among a larger fraction of the sources.

We have recently obtained optical spectroscopy of $\sim$ 75\% of our LAE sample (Zheng et al. 2010, in preparation), using the IMACS spectrograph on the Magellan 6.5m telescope, including the two LAE candidates (NB673-27 and NB673-62) which 
have S/N$_{soft} >$ 1 in CDF-S and contributed about half of the X-ray signal.
One (NB673-27) is confirmed as a low-redshift emission-line galaxy based on strong continuum flux blueward of the emission line.  The other object (NB673-62) was confirmed as a LBG at z = 4.4 with little--to--no Ly$\alpha$ flux present.
As our aim is to analyze the X-ray properties of the \lya\ emitters at z = 4.5, we excluded these 2 objects in the following stacking analysis.% interloper and keep NB673-62  %, and discuss the result without and with NB673-62. 
 The remaining 20 LAEs in the central CDF-S had net and background counts of 14 and 61, which yields a S/N = 1.4, 
with an effective exposure time of 32.7 Ms. Our Monte Carlo simulations gives a probability of P(S/N$<$1.4) = 96.03\% 
for obtaining a signal as the observed one by chance.  The stacked X-ray image of the 20 LAEs are also plotted in Figure \ref{stack}, which is less distinguishable as being above the noise.
We thus give a 3-$\sigma$ upper limit on the average flux as $\langle f_{0.5-2}\rangle < $ 1.6 $\times 10^{-17}$ \fluxunit\ for the 20 LAEs in CDF-S. Including the ECDF-S data, 
the effective exposure increases to 48.9 Ms, implying a decreased 3-$\sigma$ (1-$\sigma$) upper limit of average flux as $\langle f_{0.5-2}\rangle \le $ 1.2 $\times 10^{-17}$ \fluxunit\ (6.3 $\times 10^{-18}$ \fluxunit), corresponding to a luminosity of $\langle L_{0.5 - 2 keV}\rangle \le$  2.4 $\times 10^{42}$ ergs $s^{-1}$ (1.2 $\times 10^{42}$ ergs $s^{-1}$).

%This implied a weak signal of average flux of  
%$\langle f_{0.5-2}\rangle \sim $ 7.2 $\pm$ 3.6 $\times 10^{-18}$ \fluxunit. The stacked X-ray image of the 21 LAEs are also plotted in Figure \ref{stack}.
%Including the ECDF-S data, the average flux
%decreased to $\langle f_{0.5-2}\rangle \sim $ 4.9 $\pm$ 2.8 $\times 10^{-18}$ \fluxunit, but with much poor S/N of 1.8. 
%So we give a 1-$\sigma$ upper limit of average flux as $\langle f_{0.5-2}\rangle < $ 7.7 $\times 10^{-18}$ \fluxunit\ when 
%considering all the LAEs in the following analysis.  
%{\it So in the following analysis we only choose the average flux of the 21 LAEs in the central CDF-S region, we treat it as a marginal detection of the %stacking process. }

%we should give a 1 $\sigma$ upper limit of average flux as $\langle f_{0.5-2}\rangle < $ 7.7 $\times 10^{-18}$ \fluxunit,
%and we choose this upper limit when considering all the LAEs in the following analysis.  

%3 $\sigma$ (1 $\sigma$) 
%upper limit of average flux of $\langle f_{0.5-2}\rangle \leq$ 1.57 ($\leq$ 0.86) $\times 
%10^{-17}$ \fluxunit. Including the ECDF-S data, the  3 $\sigma$ (1 $\sigma$) upper limit 
%of average flux decreased to  $\langle f_{0.5-2}\rangle \leq$ 1.1 ($\leq$ 0.60) $\times
%10^{-17}$ \fluxunit. 

\section {DISCUSSION}

\subsection{Quasar contribution to LAEs}

One LAE (J033127.2-274247) was detected in X-ray in ECDF-S, which was
spectroscopically identified as a z = 4.48 unobscured AGN with \lya\
luminosity of $L_{Ly\alpha} = 2.4\times10^{43} \ergsec$. This yields a direct high-z \lya\ quasar density of 
2.7$^{+6.2}_{-2.2}\times$10$^{-6}$ Mpc$^{-3}$ (1 $\sigma$ Poisson error, Gehrels 1986), which is consistent with luminosity function
(XLF) of AGNs at high-redshift (the comoving space density for all
spectral type AGNs with 43 $<$ log L$_x$ $<$ 45 at redshift 4 $<$ z
$<$ 5 is 2.3$\times$10$^{-6}$ Mpc$^{-3}$, Yencho et al. 2009). 
Since \emph{Chandra} ACIS does not have uniform sensitivity across the 
field of view, it is hard to directly get the fraction of galaxies hosting an quasar
with X-ray luminosity above some value of Lx (e.g., see Figure \ref{fslya}, there are 3 LAEs with X-ray upper limit 
fluxes higher than the detected one in ECDF-S).  If we only consider the LAEs in CDF-S, then 
the type 1 quasar fraction should be $\leq$5\% with L$_{0.5-2 keV}$ $> 2\times 10^{43}\ergsec$.  

Following Wang et al. (2004) and Malhotra et al. (2003), we compare
the X-ray to \lya\ flux ratios of LAEs with three known high redshift type 2
quasars (see Figure \ref{fslya}), CDF-S 202 ($z$
= 3.7; Norman et al. 2002), CXO 52 ($z$ = 3.288; Stern et al. 2002),
and HDFX 28 ($z$=2.011; Dawson et al. 2003), and with a type 1 quasar
template derived from Sazonov et al. (2004)\footnote{$f_{ Ly\alpha }
/f_{0.5-10 keV} \sim$ 1/20 at z $\sim$ 4.5, Eq. 18 of Dijkstra \&
Wyithe 2006, and $f_{ Ly\alpha } /f_{0.5-2 keV} \sim$ 1/8 at
z $\sim$ 4.5}. The \lya\ selected AGNs at z = 4.5 (a type 1 AGN, this work),  at z = 3.1 (a type 1 AGN, Gronwall et al. 2007), and at z = 2.25 
(nine AGNs\footnote{We found that there are some mis-match in Table 5 of Nilsson et al. 2007, where they gave the X-ray detected LAE candidates. So we choose the COSMOS X-ray source catalog from Cappelluti et al. (2009) to match the LAEs of Nilsson et al. (2009). There are 9 AGNs (excluded GALEX detected) matched within a separation of 3 arcsec, and one of them is only detected in hard X-ray band. }, Nilsson et al. 2009)
are also plotted in figure \ref{fslya}. Since there are many values from different redshifts, the figure \ref{fslya} is plotted in luminosity, and the soft X-ray luminosities are converted by assuming a photon index of $\Gamma$ = 2.
We only consider the soft band observations because they are more sensitive than the total band 0.5-10 keV.
Also, high redshift AGNs have effective power law indexes
are often different, as $\Gamma=1.8$ (type 1 AGNs) or $\Gamma < 1$
(type 2 AGNs). This introduces at least 50\% difference in X-ray photometric
flux normalization in the 0.5-10 keV band, but less than 10\% in the 0.5-2 keV band (see Figure 2 of Wang et al. 2007).
So in Wang et al. 2004, who choose $\Gamma$ = 2 to get the 1-$\sigma$ upper limit of 0.5--10 keV band flux to \lya\ ratio of the z = 4.5 LAEs at LALA field, 
their type 2 AGN fraction of $<$ 4.8\% should be 2 times larger, as $<$9.6\% compared with type 2 AGN like CXO 52. 

After scaling the X-ray luminosities with \lya\ line luminosities, most of the \lya\ selected AGNs are located within the region where type 1 and type 2 quasars are located. All the 20 LAEs in CDF-S are fainter in X-rays 
than HDFX 28, our type 1 quasar and Sazonov's template, greater than 50\% and 70\% of them 
are fainter than CDF-S 202 and CXO 52. This indicates that about half of our LAEs at z $\sim$ 4.5 can be type 2 quasars like CDF-S 202.
By comparing with LAEs in ECDF-S region, we can find that only CDF-S allow us to resolve almost all of the type 1 AGN, 
as well as some kind of type 2 AGN in our LAE sample.    %A
 However, that average X-ray (1-$\sigma$ upper limit ) to \lya\ ratio is 16 
and 20 times below those of type 2 quasars like CDF-S 202 and CXO 52,
and 31,  40 and 78 times below our LAE-QSO, the  type 1 quasar template and LAE-QSO at z = 3.1. 
This implies that $<$ 6.3\% of our LAEs can be 
type 2 AGNs like CXO 52 and CDF-S 202, and $<$ 3.2\% of our LAEs can be type 1 quasar like our LAE-QSO. 

\subsection{SFR from X-ray and Escaping fraction of Lyman-$\alpha$ photon}

The average flux (3-$\sigma$ upper limit) of %marginal detection of 
our stacking analysis in the soft band corresponds to an average X-ray
luminosity of $\langle L_{0.5 - 2 keV}\rangle$ = 2.4 $\times 10^{42}$ ergs $s^{-1}$.  
 If we assume that this is due to high mass X-ray
binaries, using the empirical relation between the 0.5-2 keV
luminosity and SFR of the nearby star-forming galaxies (Ranalli et
al. 2003), we derive the upper limit of star formation rate as SFR$_X \leq$ 
%290$\pm$130
530  M$_{\sun} yr^{-1}$. This SFR is much higher than previously measured for LAEs. For comparison, Gronwall et al.'s
X-ray undetected LAEs at z=3.1 can be translated to a 3-$\sigma$ upper
limit of star formation rate of $<$ 70 M$_{\sun} yr^{-1}$, only $\sim$13\% of our upper limit 
SFR$_X$ at z=4.5. If we adopt the more recent X-ray to star formation rate calibrations
from Rosa-Gonzalez et al. (2009) and Mas-Hesse et al. (2008), based on
the XMM-Newton observation and synthetic model of starbursts,
respectively, the SFR upper limit estimated above will decrease by a factor of 1/3 $\sim$ 2/3.
Even then it is more than 30 times larger than the SFR from the \lya\ emission 
for our z $\sim$ 4.5 LAEs,which has a median of SFR$_{Ly\alpha}$ = 5.2 M$_{\sun}$ yr$^{-1}$, compared to  
$\le$ 10 times at z = 3.1 (Gronwall et al. 2007) and z = 2.1 (Guaita et al. 2010).
%%The discrepancy can be explained as the hidden AGN in our LAE sample, and 
%%dust can also enlarge the ratio between unobscured SFR to SFR$_{Ly\alpha}$. 
Our larger upper limits stem from a combination of factors--- the larger luminosity distance
at $z=4.5$; a somewhat smaller sample; and the presence of a nearly-significant
signal in our stack, which may indicate the presence of weak AGN among our sample.
%We note that the SFR$_{Ly\alpha}$ and LAE density does not change in the magnitude 
%in the redshift range of 3$\sim$6 (e.g., Figure 19 of Ouchi et al. 2008, or Figure 8 of Dawson et al. 2007).  
%So if the LAEs at z = 2.1 and 3.1 are same to LAEs at z = 4.5, the 3-$\sigma$ upper limit of SFR$_X$ = 43 M$_{\sun} yr^{-1}$
%at z = 2.1 (Guaita et al. 2010) and 70 M$_{\sun} yr^{-1}$ at z = 3.1 (Gronwall et al. 2007) can be converted to an upper limit at z = 4.5, 
%as SFR$_X(z = 4.5)$ = SFR$_X(z_0) \times$ $d_L(z = 4.5)/d_L(z_0)$ =  270  and 170  M$_{\sun} yr^{-1}$, respectively. This tells that 
%The discrepancy of SFR$_X$ at z=3.1 and z=4.5 and 
%SFR$_X$/SFR$_{Ly\alpha}>$30 show that weak AGN might be hidden in our sample. 
%% (corrected-zhenya XXXX A couple things - first, you jump from our SFR at z=4.5 to those at z=3-6 - more discussion is needed there.  Also the full factor of 30 discrepancy can be partly explained by dust as well, and that should be mentioned).

Hayes et al. (2010) reported that the average escape fraction of \lya\ photons $f_{esc, Ly\alpha}$ from star-forming galaxies at redshift
z = 2.2 is $f_{esc, Ly\alpha}$ = (5.3 $\pm$ 3.8)\% by performing a blind narrowband survey in \lya\ and H$\alpha$. % They corrected the H$\alpha$
%luminosities for dust attenuation, and multiplied by the case B Ly$\alpha$/H$\alpha$ ratio of 8.7 to get the intrinsic \lya\ luminosity function LF(Ly$\alpha_0$).  Then they got the escaping fraction of \lya\ photons $f_{esc, Ly\alpha}$ as the ratio of the  observed \lya\ luminosity function LF(Ly$\alpha_0$) to the intrinsic \lya\ luminosity function LF(Ly$\alpha_{obs}$). 
%% (corrected-zhenya XXXX I don't think you need to give all these details on Hayes' method, but I'll leave it up to you).  
Since the X-ray emission
from star-forming activities is essentially unaffected by IGM and intrinsic
dust in the galaxy at high redshift\footnote{For example, Vuong et al. 2003 showed that N$_H \sim$ 10$^{22}$ cm$^{-2}$ when A$_V \sim$ 5, 
which has little effect on X-ray photons with rest-frame energy $>$ 2 keV.}, we can choose SFR$_X$ %(from Ranalli et al. 2003) 
as the upper limit of the unobscured intrinsic SFR.  (SFR$_X$ could over-estimate the intrinsic
star formation in the case where AGN provide part of the X-ray flux.)
Then we have SFR$_X \geq$ SFR$_{intr}$ , 
and SFR$_{Ly\alpha}$ = SFR$_{intr} \times$ f$_{IGM} \times$ f$_{esc, Ly\alpha}$. 
Songaila (2004) measured the transmission of the \lya\ forest produced by IGM up to
redshift 6.3. The transmitted fraction f$_{IGM}$ is $\sim$0.3 at redshift z = 4.5, and $\sim$0.7 at z=3.1. So we can get the lower limit of 
escaping fraction of \lya\ photons as f$_{esc, Ly\alpha} \sim$ 3.2\% for z = 4.5 LAEs, and $\sim$ 9\% for z = 3.1 LAEs\footnote{$\langle SFR_{Ly\alpha}(z=3.1)\rangle$ = 4.45 M$_{\sun}$ yr$^{-1}$} with the SFR$_X$ relation from Ranalli et al. (2003). 
The lower limit of f$_{esc, Ly\alpha}$ could rise by an additional factor of 2--3 based on
the recent SFR$_X$ calibrations from Rosa-Gonzalez et al. (2009) and Mas-Hesse et al. (2008), 
which show that more X-ray photons are produced through star-forming activities. 
%% (XXXX I think that the last few sentences here need to be better explained).
 %Hayes et al. (2010) show an anti-correlation of f$_{esc}$ with dust content, and \lya\ selection preferentially finding galaxies with higher  f$_{esc}$ values. So our  f$_{esc}$ is consistence with their work.

\subsection{Existence of weak AGN in High-z Star-Forming Galaxies?}
%% XXXX I get this section, but I think it needs some work to be fully understood by the reader
%%
 Any weak AGNs at high--redshift should be captured in the narrow-band surveys, provided their Ly$\alpha$ emission is strong enough. However, 
the  AGN fractions among LAE samples reported in the literature refer
to quasar fractions (L$_x > 4\times$ $10^{43}\ergsec$) for all samples at redshifts z $>$ 3.
This is mainly due to the 
inadequate depths of X-ray exposures, apart from the two Chandra deep fields.
At z $\sim$ 2.1, Guaita et al. did not report the X-ray luminosities, 
which can be as low as 10$^{42}\ergsec$ in CDF-S.  In the local universe,  
a large fraction of weak AGNs were reported based 
on multiple methods including X-rays (Finkelstein et al. 2009a).
Although the LAEs of Gronwall et al.\ are located in the CDF-S where
the X-ray luminosity is complete above 8$\times 10^{42}\ergsec$ at z $\sim$ 3.1, 
they only found one X-ray detected LAE in ECDF-S, with L$_{X}$ = 2.8 $\times 10^{44}\ergsec$. 
They used a stacking analysis to derive a 3 $\sigma$ upper limit of  3.8 $\times 10^{41}\ergsec$ 
on the mean 0.5 - 2 keV luminosity of their LAEs.  Their stacking is based on the 
old 1-Ms CDF-S, but when we repeated this stack using the 2-Ms CDF-S data,
we also found no signal (S/N$<$1). The new 
data decreased the 3-$\sigma$ upper limit luminosity to 3.1 $\times 10^{41}\ergsec$ at z $\sim$ 3.1. 
In contrast, our $z\approx 4.5$ stacking analysis gives a 3-$\sigma$ upper limit luminosity to 
luminosity of $\langle L_{0.5 - 2 keV}\rangle$ = 2.4 $\times 10^{42}$ ergs $s^{-1}$, where weak AGNs
with luminosity of $\sim10^{42}$ ergs $s^{-1}$
might be hidden.
%This X-ray luminosity is in the range of the d

As mentioned in \S 1, AGNs could be the cause of high EWs of LAEs. 
The rest frame EW of two \lya\ selected AGNs in CDF-S (see Figure \ref{lyaew}) are $\sim$30\AA\ (our work) and $\sim$100\AA\ (Gronwall et al. 2007). 
In Subaru/XMM Deep field survey, Ouchi et al. (2008) reported two \lya\ selected AGNs with rest frame EW of 60-70 \AA\ at z = 3.1 and z =3.7. 
At z = 2.25, all the nine \lya\ selected AGNs show a rest frame EW range of 25\AA\ $<$ EW$_{rest}  <$ 160 \AA\ (Nilsson et al. 2009).
%The one LAE which shows S/N$_{soft} >$ 1 (Figure 2) also has EW$_{rest} <$200\AA (See Figure \ref{lyaew}). 
%So we divide our 22 LAEs in central CDF-S to two subsamples from their rest frame EWs. We found that when choosing 
%EW(Ly$\alpha$) $<$ 400\AA\, the subsample gets its maximum S/N at soft band as S/N = 2.7 (See Table \ref{stkdata}). 
%When decreasing the criteria to $<$200 \AA, the S/N of the subsample is also as high as 2.6.
Although the intrinsic \lya\ EW for AGN is uncertain, the rest-frame \lya\ EWs of bright AGNs are typically in the range 50--150 \AA\ 
(Charlot \& Fall 1993, and references therein). Charlot \& Fall (1993) show that AGNs which are completely 
surrounded by neutral hydrogen gas have rest-frame \lya\ EWs of 827$\alpha^{-1}(3/4)^{\alpha}$ \AA\ (ignoring absorption by dust), 
where $\alpha$ is the spectral index blueward of the \lya\ line. According to the template of Sazonov et al. (2004), $\alpha$ = 1.7, which
yields EW$_{rest}$ $\sim$ 300 \AA. Considering the scattering in the IGM, Dijkstra \& Wyithe (2006) show the intrinsic distributuion of EW should
be centred on EW$_{rest}$ = 100 \AA\ with $\sigma_{EW}$ = 30 \AA. We also check the three type 2 quasars in Figure \ref{lyaew}. 
Only type 2 quasars like CXO 52 would be selected as a LAE candidate with a large EW; the other two are either too faint or have an insufficient 
narrowband-to-broadband contrast to be selected as LAEs.
Prior to examination of the optical spectra of our LAEs, only from the view of X-ray and optical images, we found that  
LAEs in CDF-S with EW$_{rest}$(Ly$\alpha$) $<$ 400\AA\ dominate the signal as shown in Figure \ref{stack}--- indeed, this subsample has
a soft band S/N as high as 2.7 (See Table \ref{stkdata}). This is mainly due to the two LAEs which show S/N$_{soft} >$ 1 (Figure 2) and
contribute about half of the net counts.  As mentioned in Sec. 3.2, spectroscopic results show that the two LAE candidates are not \lya\ galaxies at $z\approx 4.5$. 
%We recently obtained optical spectroscopy of these two candidates, and find that neither appears to be a strong \lya\ emitting galaxy at $z\approx 4.5$.
%Rather, one is likely a foreground oxygen line emitter, and the other is likely a $z\approx 4.4$
%Lyman break galaxy without strong \lya\ emission.
%These spectra are part of a larger CDF-S spectroscopic program using the IMACS 
%spectrograph at the Magellan~I Baade telescope.  Most of the other LAE candidates in our X-ray
%stack that had spectra from the same observing run {\it were} confirmed.   
%The spectroscopic program will be described in detail elsewhere. 
%Only one (NB673-62) can be selected as a %LAE at z $\sim$ 4.5. 
Excluding these two objects from the stack, the subsample with EW$_{rest}$(Ly$\alpha$) $<$ 400\AA\ decreased to a S/N of 1.7.  
 This level of signal could be simply a Poisson fluctuation in
the photon statistics.   Alternatively, it may be due to some low-luminosity AGN in the sample (as seen
in the low-redshift \lya\ selected AGNs at $z \sim 0.3$),
or to star formation in the modest number of foreground and Lyman break galaxies that
enter the sample.  Low-luminosity AGN entering our sample could be either type 1 or type 2.
The type 1 AGN are most likely confined to the EW$_{rest}$(Ly$\alpha$) $<$ 400\AA\  subsample,
while the type 2 AGN show a larger dispersion in both EW$_{rest}$  [Fig.~\ref{lyaew}] and other
properties [Fig.~\ref{fslya}].  This can be explained by the distinct mechanisms for the extinction of \lya\
photons and the X-ray absorption for type 2 AGN, e.g., extinction of \lya\ photons by Narrow Line Region and absorption of X-ray photons by dust torus. 
Then, most \lya\ selected AGNs are likely to be hidden in the low EW$_{rest}$ region.

\section{Conclusion}
%We report the first X-ray detection of a X-ray LAE at z $>$ 4. The
Our work shows that X-ray observation is an effective method to identify AGN, 
as well as foreground objects in LAE samples.
One X-ray detected LAE is spectroscopically confirmed as a 
type 1 quasar at z = 4.5.   A stack of 22 other LAEs in the CDF-S field yields
a marginal detection.   However, two of these 22 sources contribute about
half of the stacked X-ray signal, and these two 
were found to be a foreground interloper and a LBG at z=4.4 without 
strong \lya\ emission.
The mean flux of the remaining 20 sources, while positive, is not significantly different from zero.
Including the ECDFS data,  we obtain a $3\sigma$ upper limit on the average X-ray luminosity 
 of  2.4 $\times 10^{42}$ erg s$^{-1}$. Compared to their average \lya\ luminosity, we estimate that that fewer than 3.2\% (6.3\%) 
of our LAEs could be high redshift type 1 (type 2) AGNs, and those hidden AGNs might show low EW$_{rest}$. 
Using the relationship of X-ray emission and star-forming activity from low redshift star-forming galaxies, 
we obtained an upper limit on the unobscured SFR of SFR$<$ 180-530 M$_\sun$ yr$^{-1}$. Compared to the SFR estimated from 
their average \lya\ luminosity, we find a lower limit on the escape fraction of \lya\ photons, f$_{esc,Ly\alpha} >$ 3-10\%. 
%more like LAEs at z = 4.5, and a couple of weak AGN are likely hidden. 
%  This signal 
%could be dominated by either type 1 AGN or
%star-forming galaxies. It is very interesting that at the same field,
%Gronwall et al. (2007) didn't find any significant signal with a
%stacking analysis on 160 LAEs at z=3.1. Their result is based on the
%old 1-Ms CDF-S, but when we repeated this with 2-Ms CDF-S, the signal
%remains insignificant (S/N$<$1). The new data decreased the 3 $\sigma$
%upper limit of SFR$_X$ at z=3.1 to 25\% of our average
%SFR$_X$ at z=4.5.  This indicates that our composite X-ray detection 
%is likely to be dominated by a couple of weak AGN.  
Doubling the depth of CDF-S X-ray observations is planned in 2010 and 2011 (see
\textit{Chandra} Electronic Bulletin 89).   
% We plan also to  deepen the exposure of NB656 to get more LAEs at z $\sim$ 4.5. 
This will strengthen the power of X-ray diagnostics of LAEs, 
especially for revealing their unobscured SFR, for 
the new discovery of \lya\ selected quasars and weak AGN, and for
excluding the low-redshift contamination.

%Deeper X-ray observation and more LAEs will help us to 

% and will help determine the
%true origin of this X-ray emission.

\acknowledgements 
This work is supported by National Basic Research Program of China
(973 program, Grant No. 2007CB815404), 
and Chinese National Science Foundation (Grant No. 10825312, 10773010).
The work of JER and SM is supported in part by the 
United States National Science Foundation grant AST-0808165. We thank Neal Miller for helpful discussions. We thank the referee for the 
insightful comments that helped us improve the paper significantly.

\clearpage

\begin{deluxetable}{lc|cccccc||ccc|c}
\rotate
\tablecaption{Stacking results of undetected LAEs located in the ECDF-S and
CDF-S Field with off-axis-angel $<$ 8\arcmin.}
\tablecolumns{12} \tablewidth{0pt} 
\startdata \hline\hline
X-ray Field & Number$^{a}$       &      \multicolumn{6}{c||}{X-ray COUNTS$^{b}$} & \multicolumn{3}{|c|}{F$_X$($>3\sigma$)$^{c}$} & Time \\
        &             & Net$_{S}$ & Tot$_{S}$ &  Net$_{H}$ & Tot$_{H}$ & Net$_{05-7}$ & Tot$_{05-7}$ & F$_{soft}$ &F$_{hard}$ & F$_{full}$  & Ms  \\\hline
ECDFS       & 83  &  0.2 & 26 &     -3.2 & 64 & -3.1 & 90           & 1.44 & 7.49 & 3.98 &16.2 \\
ECDFS+QSO  & 84 & 11.9  & 38  & -1.5  & 66  & 10.3 & 104 & 2.58 & 7.53 & 5.51 & 16.4  \\
\textbf{CDF-S}       & 22 &  \textbf{26.6} & \textbf{101} & 0.6 & 157 &  27.2 & 258 & 1.99 & 5.17 & 4.29 & 35.8 \\
\textbf{CDF-S$^*$}       & 20 &  \textbf{14} & \textbf{75} & 1.4 & 127 &  15.4 & 202 & 1.57 & 5.27 & 3.67 & 32.7 \\\hline
  {\scriptsize   CDF-S (EW(Ly$\alpha$)$_{rest} \ge $400\AA) } & 10 & 1.8 & 36 & 3.0 & 73 & 4.8 & 109 & 1.6 & 9.8 & 5.1 & 15.8 \\
 \textbf{{\scriptsize   CDF-S(EW(Ly$\alpha$)$_{rest}<$400\AA)}}$^d$& 12 & \textbf{24.8} & \textbf{65} &  -2.4 & 84 & 22.4 &149 & 3.1 & 6.9 & 6.1 & 20.0 \\ 
      \textbf{{\scriptsize   CDF-S(EW(Ly$\alpha$)$_{rest}<$400\AA)}}$^*$& 10 & \textbf{12.2} & \textbf{39} &  -1.6 & 54 & 10.6 & 93 & 2.4 & 6.3 & 5.0 & 16.9 \\  
    \hline
ECDFS+CDF-S$^*$ & 86 &  14.2 & 101 &  -1.8 & 191 &  12.3 & 292 & 1.17 & 3.92 & 2.71 & 48.9  \\\hline
\enddata
\tablenotetext{a}{ Number of LAEs selected for stacking analysis. 17 LAEs were covered by both CDF-S and ECDF-S, and 1 LAE in ECDF-S was covered by two ECDF-S pointings.} 
\tablenotetext{b}{Notice that the number of counts are extracted from their 50\% PSFs and summed up, 
and not corrected for the apertures.} 
\tablenotetext{c}{ The $3\sigma$ flux limitss are obtained by first calculating the 
$3 \sigma$ upper limit on counts as $Net$+3$\times(\sqrt{Tot+0.75}+1)/(\hbox{PSF-fraction})$.  Here ``Net'' and ``Tot'' are the net and total counts in the 50\% PSF region,  respectively, while ``PSF-fraction'' here is 50\%.  The counts are then divided by 
effective integration time and multiplied by the count-rate to flux convertsion factor. 
The tabulated flux limits are in units of 10$^{-17}$ \fluxunit.} 
\tablenotetext{d}{The 22 LAEs in central CDF-S are divided into two subsamples from their rest frame equivalent width. We found that when choosing EW(Ly$\alpha)_{rest} <$ 400\AA, the subsample gets its maximum S/N at soft band as S/N = 2.73. } 
\tablenotetext{*}{Two LAEs which show S/N$_{soft}>$ 1 and contribute about half of the marginal signal were excluded in the analysis. }
\label{stkdata}
 \end{deluxetable}

\begin{figure}
\plotone{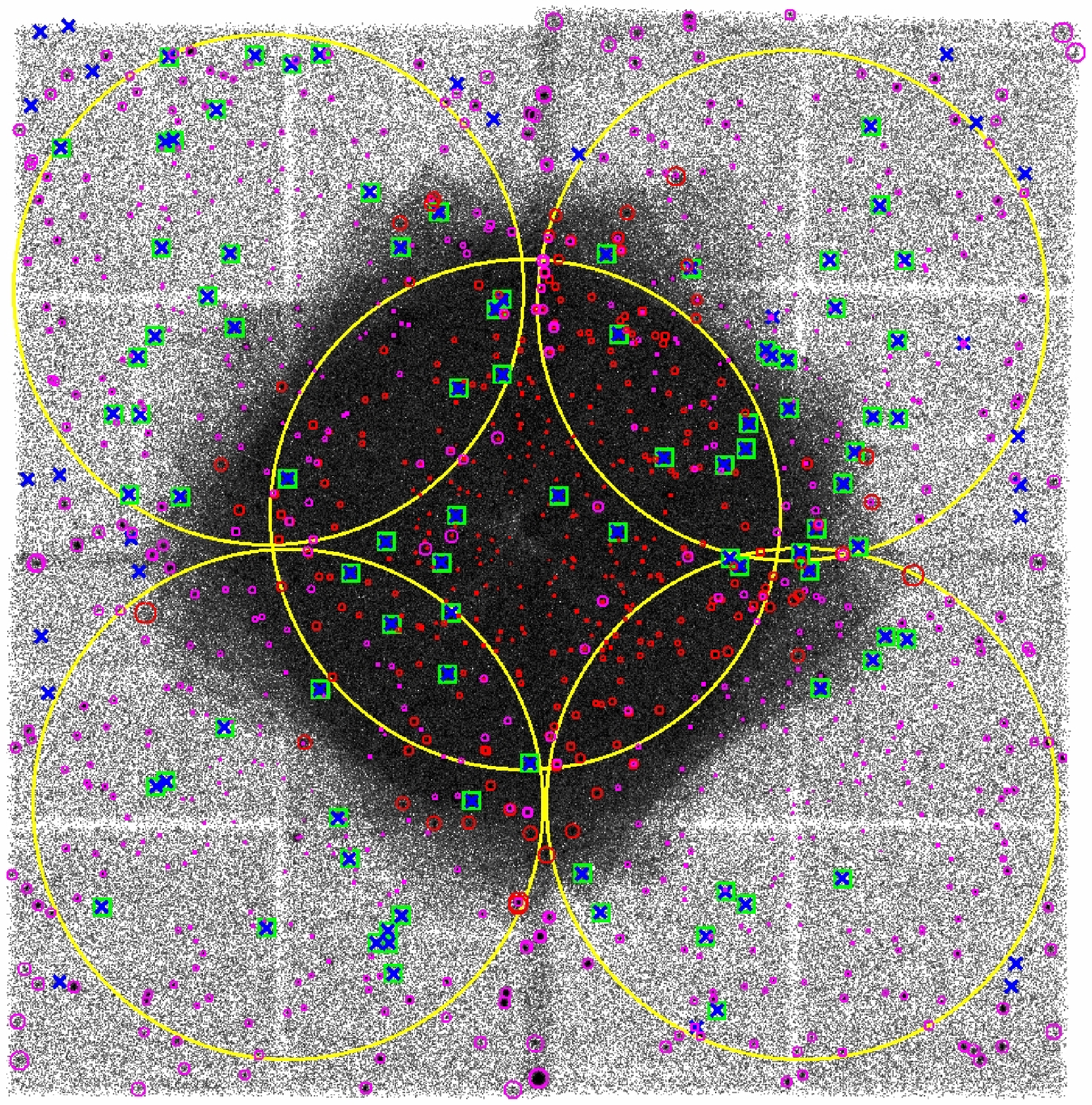} \caption{ The \textit{Chandra} image of CDF-S plus
ECDF-S.  X-ray detected sources are marked with red (CDF-S) and magenta (ECDF-S) circles, while optically selected
\lya\ emitters (LAEs) are marked with blue ``X'' s.  Those LAEs selected for 
our stacking analysis are marked by green boxes. The large yellow
circles presented the selection area (off-axis angle $\theta <$
8\arcmin) in each ACIS-I image where we considered LAEs for inclusion
in our analysis. } \label{field}
\end{figure}

\begin{figure}
\plotone{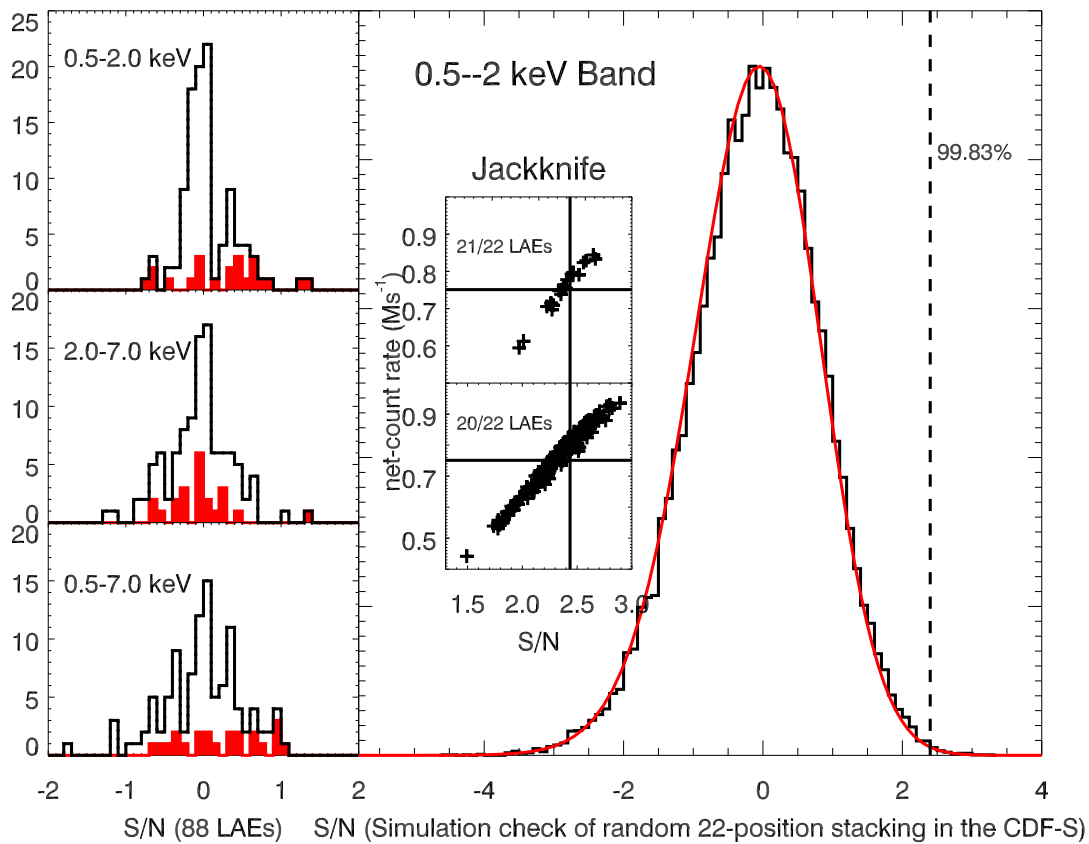} 
\caption{Histograms of the X-ray
signal-to-noise ratios (S/N) for our selected 88 \lya\ sources in three Chandra energy bands  (\textit{Left}; the 22 LAEs located in the CDF-S are filled with red colors), 
Jackknife tests on the 22 LAEs in the CDF-S at soft band (\textit{insets}), and S/N distribution of Monte Carlo simulation by using 22
random locations (fake ``sources'') in the CDF-S soft X-ray data(\textit{Right}). The smooth red curve shows
the distribution of S/N derived assuming a Poisson distribution of total counts in the aperture,
with mean 74 (as in the real data).  Both calculations yeild an $0.17\%$ chance of obtaining the
observed signal by chance.  See text for details. }
\label{stacksn}
\end{figure}

\begin{figure}
\plotone{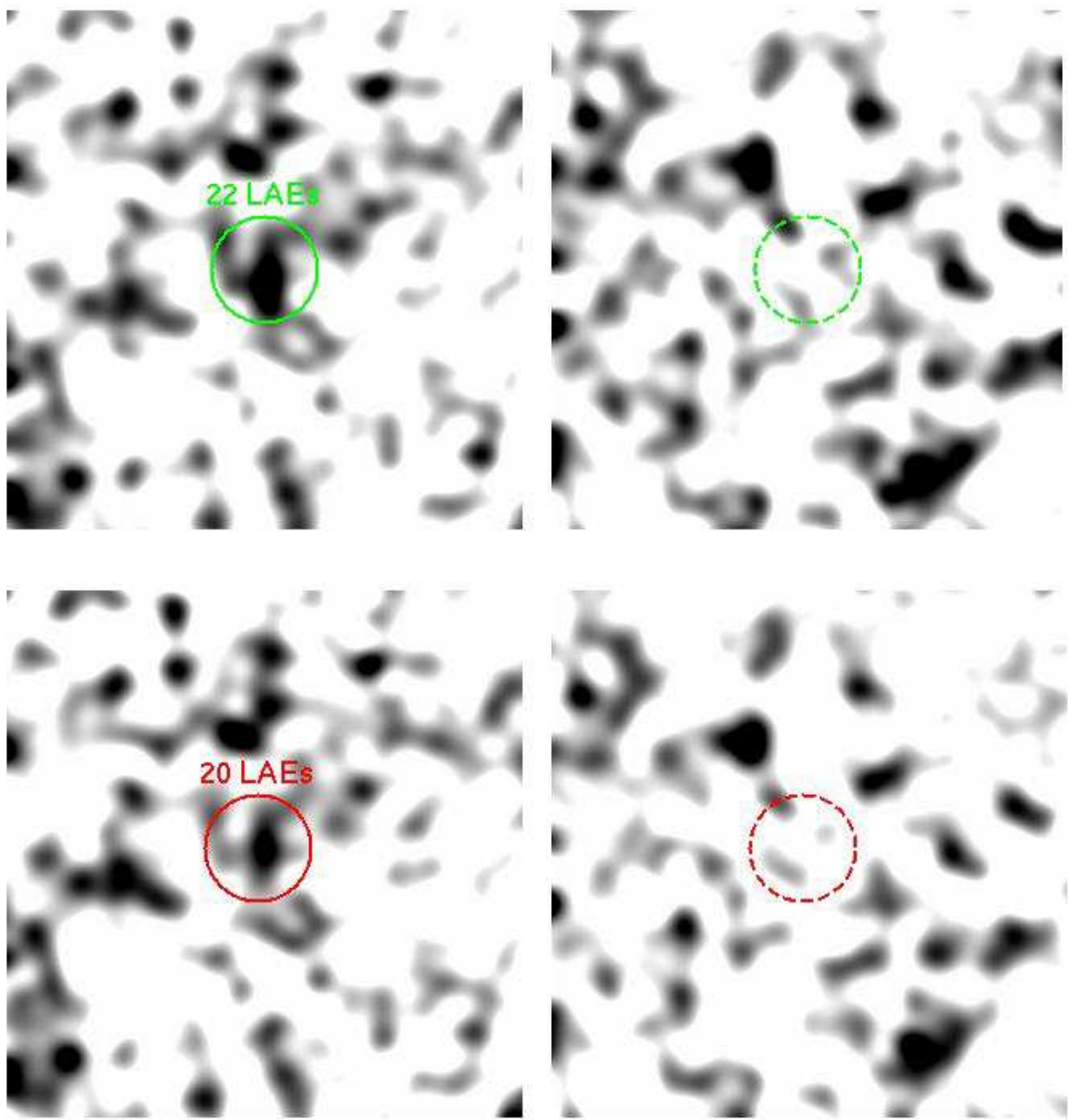} 
\caption{ The stacked X-ray image of LAEs in central CDF-S. \textit{Top}: 22 LAEs with the interloper;  \textit{Bottom}: 20 LAEs excluding 
the interloper and the LBG at z=4.4. 
% \textit{Chandra} images of 22 LAEs in CDF-S with off-axis angle $<$ 8\arcmin. 
\textit{Left}: 0.5--2.0 keV band. \textit{Right}: 2.0--7.0 keV band. The effective exposure time of the stacked
images is $\sim$36 Ms for stacking 22 LAEs. The images are $\sim 20" \times 20"$ in size, and the circles are centered on the stacking position 
and have a radius of 2$"$. The images were smoothed using a Gaussian kernel having $\hbox{FWHM} = 1.2''$,
which approximates a matched filter for point sources detection given the
$\sim 1''$ ACIS-I spatial resolution in these data.
A marginal detection (S/N = 2.4) is seen at the top left panel, while excluding the two candidate
LAEs which contribute about half of the marginal signal,  the signal is no longer convincingly
above the brightest noise peaks in the image (bottom left). } 
\label{stack}
\end{figure}

\begin{figure}
\plotone{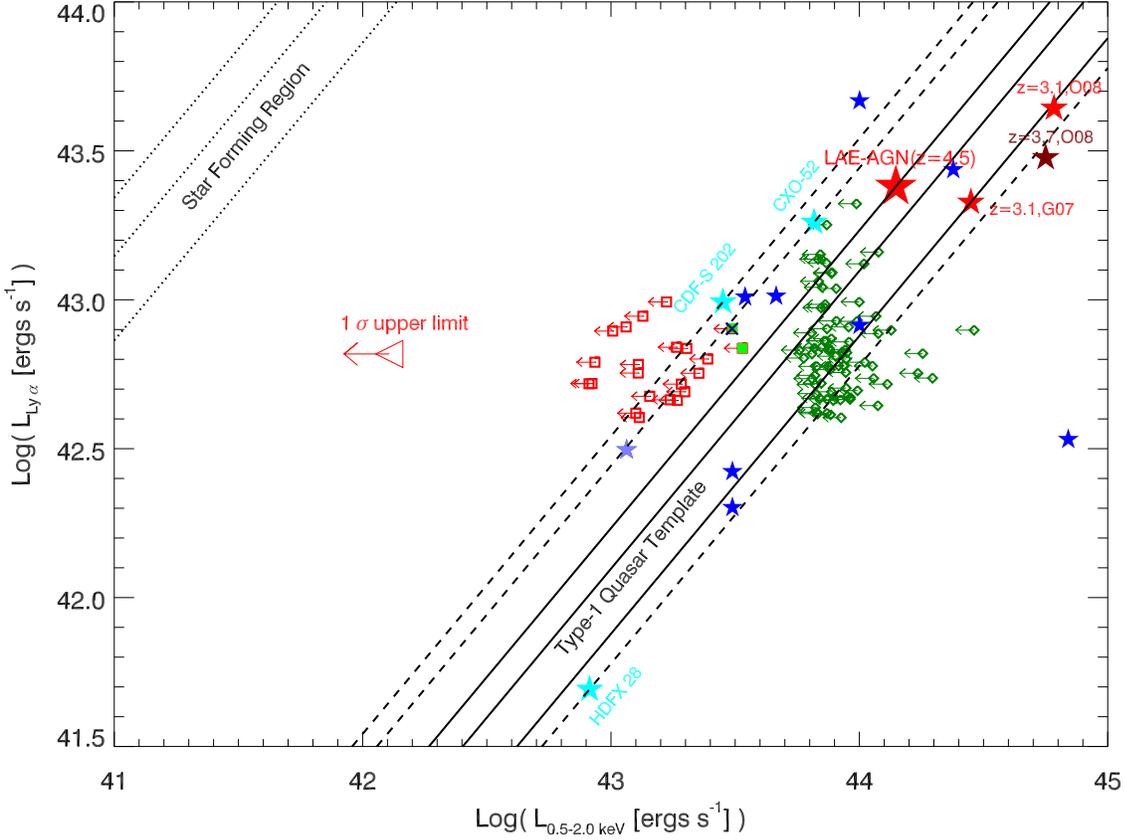} \caption{ \lya\ luminosities vs. 3 $\sigma$ upper
limits of the soft X-ray luminosities for our LAEs (red squares: LAEs in CDF-S, dark-green diamonds: LAEs in ECDF-S) at
z$\sim$4.5, compared with three known high-redshift type 2 quasars
(cyan stars). All the soft X-ray luminosities are converted by assuming photon index $\Gamma$ = 2. 
 Two LAE candidates in the central CDFS which show S/N$>$1 and 
contribute about half of the stacking signal are filled with green.  These two are excluded 
during our analysis because one (marked with a blue ``X'') appears to be a low-redshift
interloper, while the other is likely a $z\approx 4.4$ Lyman break galaxy without strong \lya\ 
emission.
Diagonal lines indicate constant X-ray to \lya\ flux ratios.
The average X-ray to \lya\ ratio for type 1 quasar template from
Sazonov et al. (2004) is plotted as solid line, and the average X-ray to \lya\ ratio for 
star-forming galaxies from Ranalli et al. (2003), Rosa-Gonzalez et al. (2009) and Mas-Hesse et al. (2008) are plotted as dotted lines.
The red stars are the \lya\ selected AGNs at z = 4.5 (this work), at z = 3.7 (Ouchi et al. 2008) and at z = 3.1(Gronwall et al 2007, Ouchi et al. 2008), 
and the blue stars are the \lya\ selected AGNs
at z = 2.25 (Nilsson et al. 2009, the one with light blue star is detected in hard X-ray band only, here the soft X-ray luminosity is the 1-$\sigma$ upper limit.)
The empty red triangle is presented as the 1-$\sigma$ upper limit from stacking at the soft band.} \label{fslya}
\end{figure}

\begin{figure}
\plotone{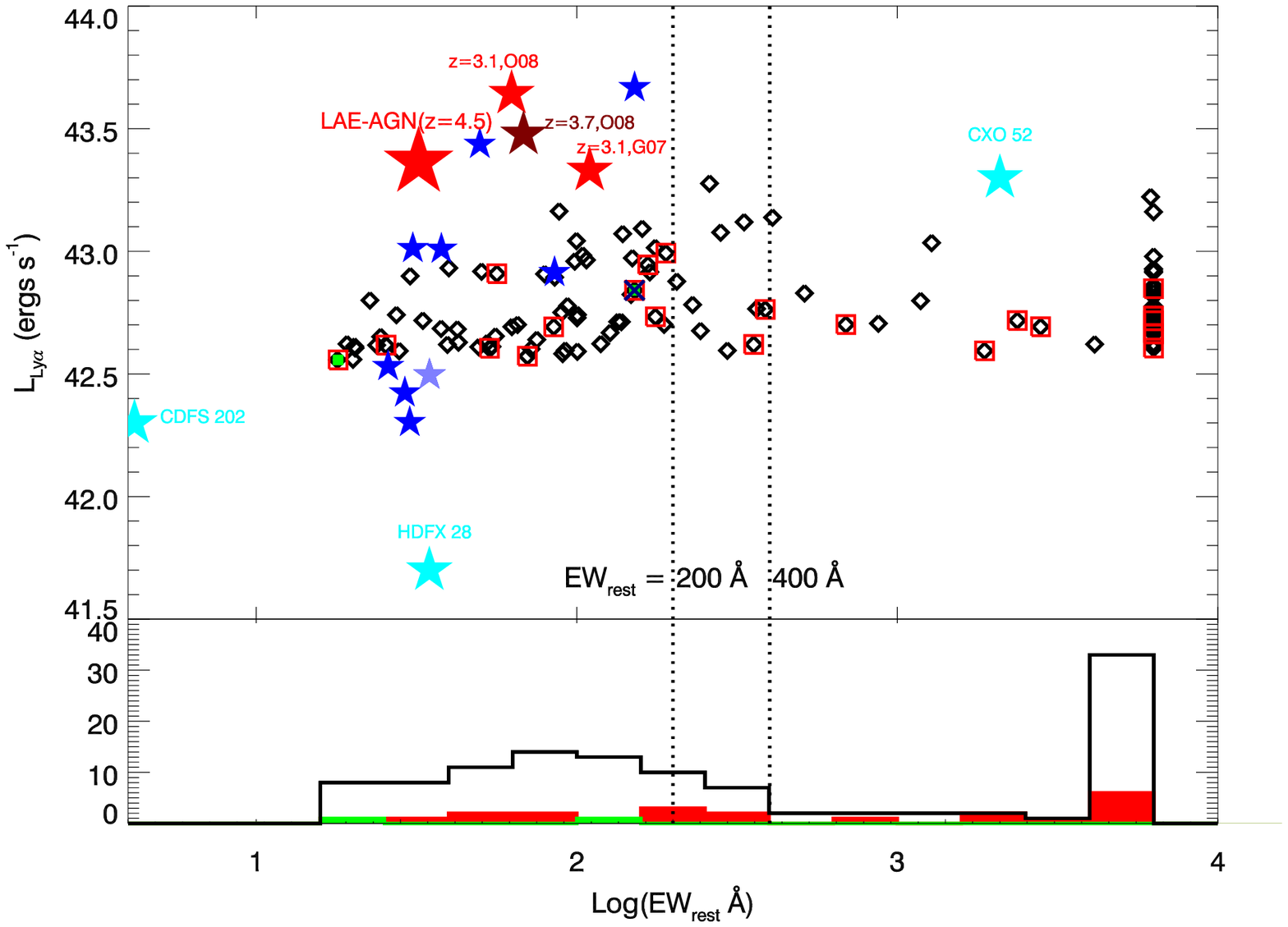} 
\caption{ Top: Rest-frame EW vs. luminosity of Ly$\alpha$ line; Bottom: Rest-frame Ly$\alpha$ EW distribution. Black diamonds: all 113 LAEs at z = 4.5; Red squares: 22 LAEs in central CDFS; Green points: 2 LAE candidates in central CDFS which show S/N$>$1 and contribute about half of the stacking signal.  The low-redshift interloper is marked with blue ``X''; the other is likely a $z\approx 4.4$ Lyman break galaxy without strong \lya\ emission.
The filled red and blue stars are one \lya\ selected AGN at z=4.5 (this work), one \lya\ selected AGN at z=3.7 (Ouchi et al. 2008), two \lya\ selected AGNs at z=3.1 (Gronwall et al. 2007 and Ouchi et al. 2008), and nine \lya\ selected AGNs
at z = 2.25 (Nilsson et al. 2009) , and X-ray selected three type 2 quasars are  marked as cyan stars. } 
\label{lyaew}
\end{figure}


\begin{references}
%\reference{} Ajiki, M., Mobasher, B., et al. 2006, ApJ, 638, 596
%\reference{} Chary, R.R., Stern, D., \& Eisenhardt, P. 2005, ApJL, 635, 5
\reference{} Charlot, S. \& Fall, M., 1993, ApJ, 415, 580		
\reference{} Colbert, E. M., et al. 2004, ApJ, 602, 231					
\reference{} Cowie, L. L., Barger, A.J., \& Hu, E. M. 2010, ApJ, 711, 928  		
\reference{} Dawson, S., McCrady, N., et al. 2003, AJ, 125, 1236					
\reference{} Dawson, S., Rhoads, J. E., Malhotra, S., et al. 2004, ApJ,  617, 707         
\reference{} Dawson, S., Rhoads, J. E., Malhotra, S., et al. 2007, ApJ, 671, 1227        
\reference{} Dickey, J. \& Lockman, F., 1990, ARA\&A, 28, 215 						
\reference{} Dijkstra, M., \& Wyithe, J.S.B., 2006, MNRAS, 372, 1575 					
\reference{} Finkelstein, S. L., Cohen, S. H., Malhotra, S., Rhoads, J. E., et al. 2009a, ApJL, 703, 162	
\reference{} Finkelstein, S. L., Cohen, S. H., Malhotra, S., \& Rhoads, J. E. 2009b, ApJ, 700, 276 		
\reference{} Finkelstein, S. L., Rhoads, J. E., Malhotra, S., \& Grogin, N. 2009c, ApJ, 691, 465 			
\reference{} Finkelstein, S. L., Rhoads, J. E., Malhotra, S., Grogin, N., \& Wang, J. X. 2008, ApJ, 678, 655 
\reference{} Gawiser, E., Francke, H., Lai, K., et al., 2007, ApJ, 671, 278 			
%\reference{} Geach, J. E., Alexander, D. M., et al. 2009, ApJ, 700, 1
\reference{} Gehrels, N. 1986, ApJ, 303, 336 									
\reference{} Grimm, H.J., Gilfanov, M., \& Sunyaev, R., 2003, MNRAS, 339, 793  			
\reference{} Gronwall, C. et al. 2007, ApJ, 667, 79 									
\reference{} Guaita, L., Gawiser, E., et al. 2010, ApJ, 714, 255							
\reference{} Hansen, M. \& Oh, S. P. 2006, MNRAS, 367, 979							
\reference{} Hayes, M., Ostlin, G., et al. 2010, Nature, 464, 562	                 
\reference{} Hu, E. M., Cowie, L. L., et al. 2004, AJ, 127, 563                 
\reference{} Iye, M., Ota, K, Kashikawa, N., et al. 2006, Nature, 443, 14 			
%\reference{} Kennicutt, R. C., Jr. 1998, ARA\&A, 36, 189 
%\reference{} Kudritzki, R. P., Mendez, R. H., et al. 2000, ApJ, 536, 19 
\reference{} Lai, K., Huang, J. S., et al. 2007, ApJ, 655, 704 						
\reference{} Laird, E.S., Nandra, K., et al. 2005, MNRAS, 359, 47					
\reference{} Laird, E.S., Nandra, K., Hobbs, A., \& Steidel, C.C. 2006, MNRAS, 373, 217 	
%\reference{} Laursen, P., Sommer-Larsen, J., \& Andersen, A. C. 2009, ApJ, 704, 1640
 \reference{} Lehmer, B.D., Brandt, W. N., et al. 2005, ApJS, 161, 21 				
%\reference{} Lehmer, B.D., Brandt, W. N., et al. 2008, ApJ, 681, 1163 
\reference{} Lehmer, B.D., Alexander, D.M., et al. 2009, ApJ, 691, 687 			
\reference{} Luo, B., Bauer, F. E., Brandt,W. N., et al. 2008, ApJS, 179, 19					
\reference{} Malhotra, S. \& Rhoads, J.E. 2002, ApJL, 565, 71 					
\reference{} Malhotra, S., Wang, J, Rhoads, J. E., et al. 2003, ApJ, 585, L25		
\reference{} Mas-Hesse, J. M., Oti-Floranes, H., \& Cervino, M. 2008, A\&A, 483, 71		
\reference{} Nilsson, K. K., Tapken, C., et al. 2009, A\&A, 498, 13  				 
\reference{} Neufeld, D. A. 1991, ApJL, 370, 85							
\reference{} Norman, C. et al. 2002, \apj, 571, 218							
\reference{} Ouchi, M., Shimasaku, K., et al. 2008, ApJS, 176, 301      
\reference{} Persic, M., Rephaeli, Y. et al. 2004, A\&A, 419, 849				
\reference{} Pirzkal, N., Malhotra, S., Rhoads, J. E., Xu, C. 2007, ApJ, 667, 49 		
\reference{} Ranalli, P., Comastri, A., \& Setti, G. 2003, A\&A, 399, 39 			
\reference{} Rhoads, J. E., Malhotra, S., et al. 2000, ApJL, 545 85  
\reference{} Rhoads, J. E. et al., 2003, AJ, 25, 1006                         
\reference{} Rosa-Gonzalez, D., et al. 2009, MNRAS, 399, 487 					
\reference{} Sazonov, S.Y., Ostriker, J.P., \& Sunyaev, R.A., 2004, MNRAS, 347, 144		
\reference{} Scarlata, C. et al. 2009, ApJL, 704, 98							
\reference{} Songaila, A. 2004, AJ, 127, 2598 					
\reference{} Stern, D. et al. 2002, ApJ, 568, 71								
%\reference{} Taniguchi, Y., et al. 2005, PASJ, 57, 165 
\reference{} Treister, E., Virani, S., Gawiser, E., et al. 2009, ApJ, 693, 1713 					
\reference{} Venemans, B. P., et al. 2005, A\&A, 431, 793 					
\reference{} Wang, J. X., Rhoads, J. E., Malhotra, S., et al. 2004, ApJL, 608, 21 	
\reference{} Wang, J. X., Malhotra, S., \& Rhoads, J. E., 2005, ApJL, 622, 77              
\reference{} Wang, J. X., Zheng, Z. Y., Malhotra, S., et al. 2007, ApJ, 669, 765		
\reference{} Wang, J. X., Malhotra, S., Rhoads, J. E., et al. 2009, ApJ, 706, 762 	
%\reference{} Wright, S. A. et al. 2010, astro-ph/1001.5041
%\reference{} Weidinger, M., Moller, P., Fynbo, J.P.U., \& Thomsen, B. 2005, A\&A, 436, 825 
\reference{} Yencho, B., Barger, A., et al. 2009, ApJ, 698, 380					
\end{references}
\end{document}